# DISCOVERING M-DWARF COMPANIONS WITH STEPS

Steven H. Pravdo
Jet Propulsion Laboratory, California Institute of Technology
306-431, 4800 Oak Grove Drive, Pasadena, CA 91109; spravdo@jpl.nasa.gov

Stuart B. Shaklan
Jet Propulsion Laboratory, California Institute of Technology
301-486, 4800 Oak Grove Drive, Pasadena, CA 91109, shaklan@huey.jpl.nasa.gov

James Lloyd
Department of Astronomy, Cornell University
610 Space Sciences Building, Ithaca, NY 14853-6801, jpl@astro.cornell.edu
&

G. Fritz Benedict
MacDonald Observatory, University of Texas at Austin
Austin, Texas 78712-1083, fritz@astro.as.utexas.edu

## ABSTRACT

The Stellar Planet Survey (STEPS) is an ongoing astrometric search for giant planets and brown dwarfs around a sample of ~30 M-dwarfs. We have discovered several low-mass companions by measuring the motion of our target stars relative to their reference frames. We describe the STEPS method for stellar centroiding that enables our ~1 milliarcsecond relative astrometry. We describe one of the astrometrically discovered companions, GJ 1210B, which was later confirmed with an imaging observation. GJ 1210 A and B have spectral types M4-5, based on their luminosities, inferred masses, and colors.

## 1. INTRODUCTION

More than 130 extrasolar planets have been discovered (e.g. Marcy et al. 2003) in the ~10 years since the first discovery (Mayor & Queloz 1995). Most of the stars in these systems are solar-like. Few of the systems contain M-dwarfs with a notable exception of GJ 876 (Marcy et al. 2001), a M4 primary with at least two planets. The STEPS program targets M-dwarfs in an astrometric search for companions. There may exist a bounty of detectable planets around M-dwarfs, because of the preeminent position of M-dwarfs in the stellar census, and because their lower masses give rise to larger astrometric signals for a given companion mass. STEPS is a CCD-based imaging system mounted at the Cassegrain focus of the Palomar 5-m telescope that has observed its targets since 1997.

## 2. RELATIVE ASTROMETRY

The STEPS field is 2' on a side with 78 *mas* pixels. The positions are determined in 10-20 consecutive frames and combined to yield a nightly position. A target may be measured ~5 times throughout each year, often in consecutive nights over several runs.

We extract 5" image boxes containing the target and references stars after the frames are flat-fielded and cleaned. We use the image of the target star as a template for the cross-correlation centroiding in two dimensions, X and Y, separately. First we integrate the stellar images in one dimension. Then we compute the 1-D fast-Fourier transform (FFT) of the image and estimate the slope at the origin from the first value of the FFT, i.e., the amplitude for 1 cycle per box. The slope at the origin of the FFT is mathematically equivalent to the centroid, and our first-frequency value is a good approximation to the slope. (Most of the signal is in the first frequency. Higher frequencies can be used, in principle, but we have not needed to do so.) Finally, the slope differences between the reference and target stars yield the relative centroid positions when added to the nearest-integral number of pixels separating the stars. Figure 1 illustrates this process. Note that a constant background bias gives a delta-function at the origin but does not change the slopes.

We have tested other centroiding schemes and find this to be the best for several reasons. First, as mentioned above, it is insensitive to uniform background bias. Second, the target star provides the matched filter. This is crucial because the telescope suffers from noticeable time-variable aberrations including focus, astigmatism, and coma. While this limits the effectiveness of model-based (e.g. Gaussian) fits, it does not seriously degrade the matched filter approach because the variability is common to all stars in the field. Third, resampling is not required since the pixel size, 78 *mas*, is far below the typical image size of ~1". And finally, a practical consideration is that current computers efficiently process 60-point 1-D FFTs.

### 3. GJ 1210

GJ 1210 (=LHS 3265, G 139-12, Table 1) is a STEPS target. Our 6-y baseline of observations yields parallax and proper motion values (Table 2) consistent with the literature, and shows an additional astrometric motion indicative of a companion. A Monte Carlo analysis shows that the astrometric orbit alone is not well constrained due to the low absolute signal and a period that may be close to the current observing baseline.

We use imaging and spectroscopic observations to further constrain the system. There are often two distinct companion-mass solutions (Pravdo et al. 2004). To distinguish between the two we performed *H*-band adaptive optics (AO) observations with the Palomar 5-m system (Troy et al. 2000) on June 7, 2004 UT. Good seeing and low bandwidth correction with a 100 Hz update rate achieved a 10-20% Strehl ratio and allowed natural guide star observations of this relatively faint target. The observation shows that GJ 1210 is a close-binary system (Fig. 2) with a component *H*-band luminosity ratio = 0.91±0.05. We combined this luminosity ratio with the 2MASS total *H*-band result, the parallax, and the *H* mass-luminosity-relationship (MLR, Henry & McCarthy 1993) to derive a mass ratio of 0.96±0.02. We then use the *V*-band MLR (Henry et al. 1999) to constrain the *V*-band luminosity ratio to 0.74-0.92. Finally, on July 9, 2004 UT we searched for a double-lined spectroscopic binary with the Sandiford Cass Echelle spectrograph on the McDonald Observatory 2.1-m telescope (Benedict *et. al* 2001). There was a null result with an upper limit of 3 km s$^{-1}$ relative velocity. This further constrains the astrometric orbits to ones that are nearly face on or with low radial velocity at this epoch. The system mass is 0.27±0.03 M☉ with the uncertainty due to the parallax and the MLR. The spectral types for GJ 1210A and B are estimated from similar stars to be M4-5 (Kirkpatrick et al.1994). The results of >150000 Monte Carlo orbital fits to the STEPS data, after the application of the above constraints, are listed in Table 2 and illustrated in Fig. 3.

**Table 1: GJ 1210 Known Parameters**

| QUANTITY | VALUE | REFERENCE |
|---|---|---|
| RA (2000) | 17h07m40.85s | Bakos et al 2002 |
| Decl. (2000) | +07d22m04.8s | " |
| Parallax (*mas*) | 78 ± 5.3 | van Altena 1995 |
| Proper Motion magnitude (*mas*) | 660 | " |
| Proper Motion position angle (deg) | 230.83 | " |
| $V$ magnitude | 14.02 | Weis 1996 |
| $J$ magnitude | 9.284 ± 0.023 | 2MASS |
| $H$ magnitude | 8.654 ± 0.025 | " |
| $K_s$ magnitude | 8.419 ± 0.023 | " |

**Table 2: GJ 1210 STEPS Values**

| QUANTITY | VALUE |
|---|---|
| Parallax (*mas*) | 81 ± 3 |
| Proper Motion magnitude (*mas*) | 632 ± 3 |
| Proper Motion position angle (deg) | 233 ± 1 |
| Period (y) | ~1, 4-15 |
| Total Mass (M☉) | 0.27±0.03 |
| Inclination (deg) | 100-150 or 180 |
| GJ 1210A mass from optical, IR (M☉) | 0.137, 0.130 |
| GJ 1210B mass from optical, IR (M☉) | 0.132, 0.125 |

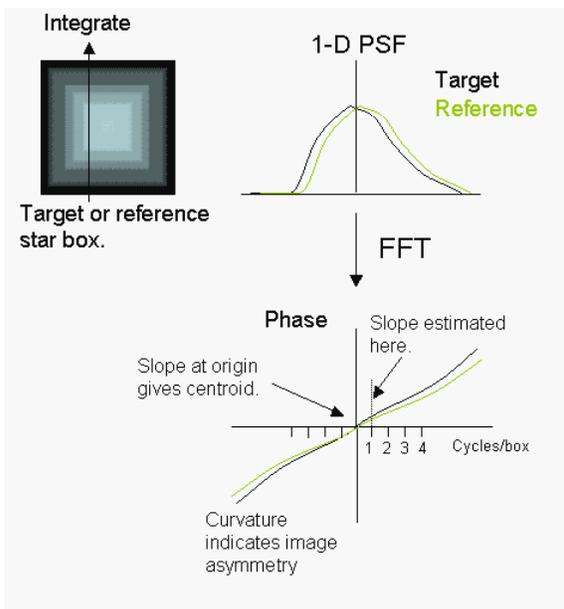

**Figure 1. STEPS centroiding method for relative astrometry.**

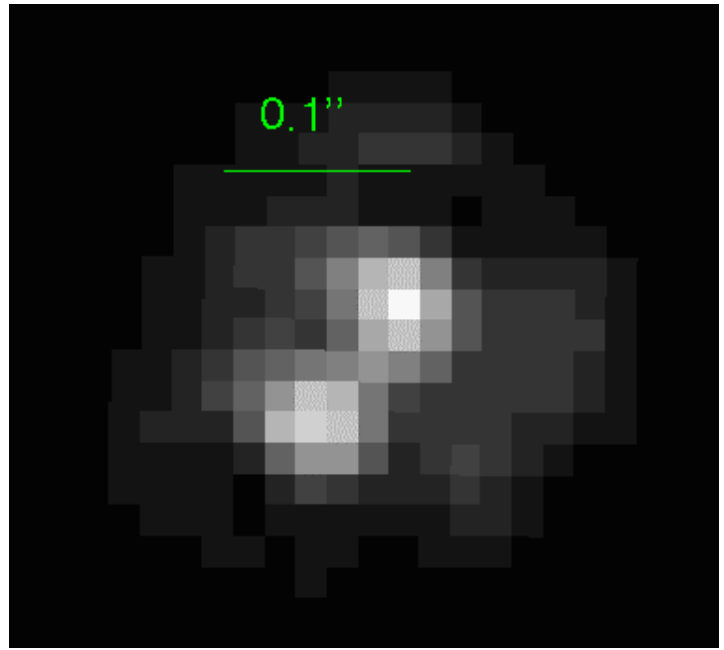

**Figure 2. AO *H*-band image of GJ 1210.**

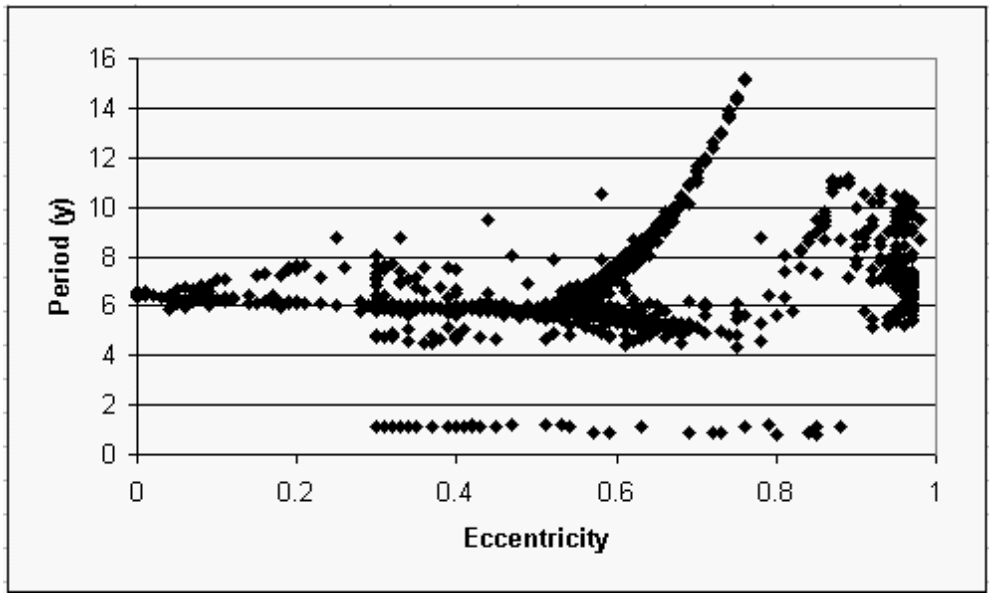

**Figure 3. Possible GJ 1210 AB orbits**

4